\documentclass[runningheads]{llncs}

\usepackage{url}
\usepackage[hidelinks]{hyperref}
\usepackage{subcaption}
\usepackage{tikz}
\usepackage{xcolor}
\usepackage{graphicx}
\usepackage{comment}
\usepackage{booktabs}
\usepackage{float}

\usepackage{csquotes}
\usepackage{enumitem}
\usepackage{siunitx}

\usepackage[acronym,shortcuts]{glossaries}
\newacronym{APT}{APT}{Advanced Persistent Threat}
\newacronym{CPS}{CPS}{Cyber-Physical System}
\newacronym{DFD}{DFD}{Data Flow Diagram}
\newacronym{DoF}{DoF}{Degree of Freedom}
\newacronym{ENISA}{ENISA}{European Union Agency for Cyber Security}
\newacronym{EUROMAP}{EUROMAP}{European Plastics and Rubber Machinery}
\newacronym{FID}{FID}{Frechet Inception Distance}
\newacronym{HiL}{HiL}{Hardware-in-the-Loop}
\newacronym{ICS}{ICS}{Industrial Control System}
\newacronym{IIoT}{IIoT}{Industrial Internet of Things}
\newacronym{IDS}{IDS}{Intrusion Detection System}
\newacronym{isia}{JRC ISIA}{Josef Ressel Center on Intelligent and Secure Industrial Automation}
\newacronym{IT}{IT}{Information Technology}
\newacronym{LQR}{LQR}{Linear-Quadratic Regulator}
\newacronym{LSTM}{LSTM}{Long Short-Term Memory}
\newacronym{MAE}{MAE}{Mean Absolute Error}
\newacronym{MSE}{MSE}{Mean Squared Error}
\newacronym{RMSE}{RMSE}{Root Mean Squared Error}
\newacronym{NC}{NC}{Numerical Control}
\newacronym{NIST}{NIST}{US National Institute of Standards and Technology}
\newacronym{NN}{NN}{Neural Network}
\newacronym[\glslongpluralkey={Operational Technologies}]{OT}{OT}{Operational Technology}
\newacronym{opcua}{OPC~UA}{Open Platform Communications Unified Architecture}
\newacronym{OSI}{OSI}{Open Systems Interconnection}
\newacronym{PCB}{PCB}{Printed Circuit Board}
\newacronym{PKI}{PKI}{Public Key Infrastructure}
\newacronym{PLC}{PLC}{Programmable Logic Controller}
\newacronym{rami}{RAMI 4.0}{Reference Architectural Model Industrie 4.0}
\newacronym{SCADA}{SCADA}{Supervisory Control and Data Acquisition}
\newacronym{SUAS}{SUAS}{Salzburg University of Applied Sciences}

\renewenvironment{abstract}{%
	\list{}{\advance\topsep by0.35cm\relax\small
		\leftmargin=1cm
		\itemindent\listparindent
		\rightmargin\leftmargin}\item[]}
{\endlist}
\usepackage[capitalise]{cleveref}
\Crefname{figure}{Fig.}{Figs.}% {<type>}{<singular>}{<plural>}
\begin{document}

%\title{A Generative Model Based Honeypot for Industrial OPC~UA Communication}
\title{A Generative Model Based Honeypot for Industrial OPC~UA Communication\thanks{Olaf~Sassnick, Georg~Schäfer, Thomas~Rosenstatter and Stefan~Huber are supported by the Christian Doppler Research Association.  
	This preprint has not undergone peer review or any post-submission improvements or corrections. The Version of Record of this contribution is accepted and will be published
	in Computer Aided Systems Theory -- EUROCAST 2024.} }

\titlerunning{A Generative Model Based Honeypot}
\author{Olaf~Sassnick\inst{1,2,3}\and %\orcidID{0000-0002-3090-4399} \and
	Georg~Schäfer\inst{1,2,3}\and %\orcidID{0009-0004-7622-7401 } \and
	Thomas~Rosenstatter\inst{1,2}\and %\orcidID{0000-0002-9587-3423} \and
	Stefan~Huber\inst{1,2}% \orcidID{0000-0002-8871-5814}}
}
\authorrunning{O. Sassnick et al.}

\institute{
Josef Ressel Centre for Intelligent and Secure Industrial Automation \and
Salzburg University of Applied Sciences, Salzburg, Austria \and
Paris Lodron University of Salzburg, Salzburg, Austria
\email{olaf.sassnick@fh-salzburg.ac.at}}
%\url{http://www.springer.com/gp/computer-science/lncs}

%
\maketitle              % typeset the header of the contribution
\begin{abstract}
Industrial Operational Technology (OT) systems are increasingly targeted by cyber-attacks due to their integration with Information Technology (IT) systems in the Industry 4.0 era. 
Besides intrusion detection systems, honeypots can effectively detect these attacks. However, creating realistic honeypots for brownfield systems is particularly challenging.
This paper introduces a generative model-based honeypot designed to mimic industrial OPC UA communication.
Utilizing a Long Short-Term Memory (LSTM) network, the honeypot learns the characteristics of a highly dynamic mechatronic system from recorded state space trajectories.
Our contributions are twofold: first, we present a proof-of-concept for a honeypot based on generative machine-learning models, and second, we publish a dataset for a cyclic industrial process.
The results demonstrate that a generative model-based honeypot can feasibly replicate a cyclic industrial process via OPC UA communication.
In the short-term, the generative model indicates a stable and plausible trajectory generation, while deviations occur over extended periods.
The proposed honeypot implementation operates efficiently on constrained hardware, requiring low computational resources.
Future work will focus on improving model accuracy, interaction capabilities, and extending the dataset for broader applications.

\end{abstract}
\section{Introduction}

\Acp{OT} in industrial production environments are
increasingly targeted by cyber attacks \cite{makrakis_industrial_2021}. In the past, \ac{OT} systems remained
completely isolated from public networks and were specifically designed for
production processes. This gradually changed with the Industry 4.0 era to
enable data-based decisions, improving the efficiency of the overall production
process. Two design principles for modern Industry 4.0 applications are the
interconnection and information transparency~\cite{hermann_design_2016},
resulting in a convergence of \ac{OT} and traditional \ac{IT}
\cite{chemudupati_convergence_2012}. The increased interconnection between
devices and systems, however, results in more sophisticated software and
hardware technologies. As such, the vulnerability against cyber attacks
increases  in a sector where availability and reliability is of utmost
importance. Consequently, advanced network security measures are required to
keep the risks of a cyber attack at a manageable level.

In this context, a honeypot can serve as a decoy and warning system.
The focus of this work is put on the creation of a honeypot that authentically mimics an actual industrial process.
Regarding the industrial process \cite{groover_automation_2008}, one can distinguish between
continuous industrial processes and discrete automation processes.
For example, a chemical plant implements a continuous industrial process, where a product is produced as a continuous stream.
In contrast, a discrete automation process is typically characterized by the assembly of units or products via a sequence of discrete steps.
Most existing datasets and testbeds for security research in the OT domain focus on
continuous industrial processes~\cite{conti_survey_2021}.
For this work our focus is on a discrete automation process with fast dynamics, like
it is given for a piece-wise production of goods.

Most publications on honeypots for \acp{CPS} rely on parametrized simulations.
Franco~\textit{et al.}~\cite{franco_survey_2021} found in their survey on honeypots in 2021
only one single approach~\cite{qiu_dipot:_2018} that employed machine-learning algorithms
out of the 44 \ac{CPS}-related studies reviewed.
Setting up a honeypot for a brownfield \ac{OT} installation by means of a simulation, however, is a challenging task,
as physical system characteristics and corresponding parameters need to be determined beforehand.

The recent advances in machine-learning enable us to replace conventional simulations with self-learning models, 
potentially resulting in a more widespread use.
In this work, we explore a generative model approach utilizing a \ac{LSTM} network to learn the characteristics 
of a highly dynamic mechatronic system from recorded state space trajectories. 
Consequently it can mimic the mechatronic system via an \ac{opcua} interface.
\ac{opcua} is used, because it is an established communication standard in the automation industry, 
enabling different manufacturers to operate seamlessly together.

The contribution of this work is two-fold:
Firstly, we propose a proof-of-concept for a honeypot of a \ac{CPS}
based on generative machine-learning models with \ac{opcua} communication.
Secondly, we introduce and publish a dataset for a highly-dynamic
cyclic process, which can be used to train generative models.

The remaining work is organized as follows:
In \cref{sec:approach} the concept of a honeypot and its deployment variants are briefly
introduced and research questions are presented.
Consequently the \ac{CPS}, which is being studied is introduced in \cref{sec:cps}, at first the physical hardware,
and secondly the its cyber representation, namely the \ac{opcua} information model.
In the following \cref{sec:data} the recorded data from the \ac{CPS} is described,
which represents a cyclic procedure, continuously carried out by the \ac{CPS}.
With the data, the generative model is trained, and an overview on the employed network 
structure is therefore given in \cref{sec:model}.
Finally, the results of the generative model are discussed in \cref{sec:results} and conclusions are given in \cref{sec:conclusion}.

\begin{figure}[ht]
	\centering
	\includegraphics[width=\textwidth]{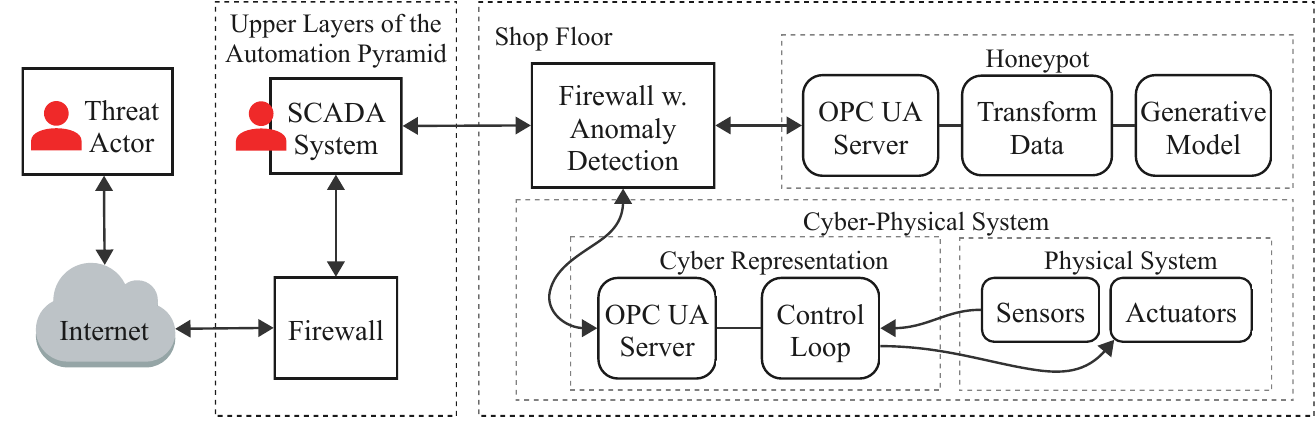}
	\caption{Cyber-physical system architecture with a honeypot for security, where the threat actor gained access to the SCADA system.} \label{fig:dfd}
\end{figure}

\section{Approach}\label{sec:approach}
In our approach, we assume a scenario as shown in \cref{fig:dfd}. 
A threat actor is attacking a production environment remotely via the Internet.
By exploiting a system vulnerability or through social engineering tactics, the threat actor gains control of the \ac{SCADA} system.
Consequently, the intruder can interact with the \ac{CPS} located on the shop floor, communicating via the firewall.
The \ac{CPS} itself implements an \ac{opcua} server and a control loop, 
while it has sensors and actuators interacting with the environment on a physical level.
The honeypot is able to replicate the \ac{opcua} server of the \ac{CPS}, 
including all relevant variables and their trajectories over time.
For the deployment of the honeypot, we propose two variants:
\begin{description}
	\item[Variant 1: permanently.] In the traditional way, matching its original definition, 
		the honeypot can be deployed permanently. As such, it must lure the attacker into interaction and 
		therefore is placed easily noticeable for the attacker.
		For example, it could communicate only with a reduced set of security measures.
		Any kind of interaction with the honeypot is suspicious and will trigger an alarm.
	\item[Variant 2: on-demand.]
	    For the second variant, the honeypot is only created once an intruder has been detected.
		At first the intruder interacts with the actual \ac{CPS}, however is being detected by the \ac{IDS} based on introduced anomalies.
		Instead of ceasing the communication channel, the intruder is forwarded to an on-demand honeypot,
		which is set up with the initial state of the \ac{CPS}, isolated from the network in the production environment.
		Consequently, all actions and observations by the intruder are performed on the honeypot instead of the real \ac{CPS}.
		As such, the interaction with the honeypot exposes information regarding the attackers behavior,
		which in return can be utilized to strengthen the system's security.

\end{description}

While \textbf{variant 1} mainly extends the available intrusion detection capabilities,
\textbf{variant 2} can be considered as a form of an automated threat detection and response.

The remaining work is guided by the following three research questions:
\begin{enumerate}[label=\textbf{RQ\arabic*}, leftmargin=1.3cm]
    \item Can a generative model-based industrial honeypot be realized by replicating a cycling process via \ac{opcua} communication?\label{rq:feasibility}
    \item What long-term characteristics can be expected from a generative model?\label{rq:characteristics}
    \item What hardware resources are required for the deployment of a generative-model-based industrial honeypot?\label{rq:resources}
\end{enumerate}

To address \ref{rq:feasibility}, we first introduce an appropriate \ac{CPS} in \cref{sec:cps}, and subsequently record a corresponding dataset (\cref{sec:data}).
The next step involves presenting a proof-of-concept for a generative model in \cref{sec:model}, 
capable of replicating the dataset's trajectories.
Following this, we discuss the necessary implementation steps to deploy 
the generative model integrated with an \ac{opcua} server, thereby completing the honeypot.
To answer \ref{rq:characteristics}, we examine the long-term characteristics in \cref{sec:results} 
by evaluating the \ac{RMSE} for multiple generated trajectories over an extended period.
Additionally, \ref{rq:resources} is addressed in \cref{sec:results}, by providing 
the specifications of the deployed system hardware and presenting runtime performance metrics for comparison.

\section{Cyber-physical System}\label{sec:cps}

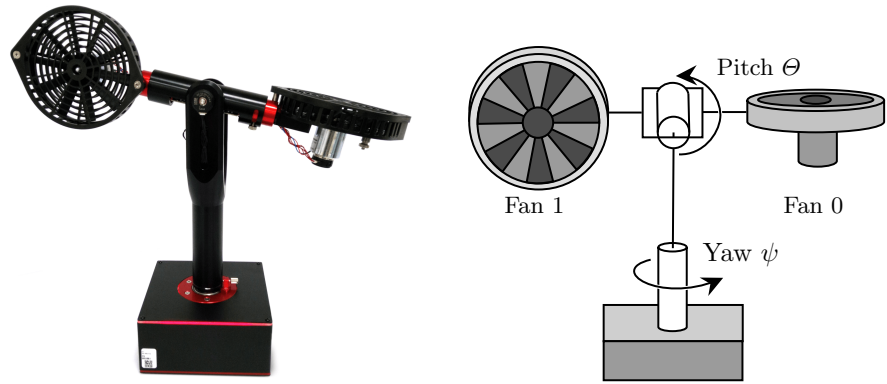
\begin{figure}[ht]
	\centering
	\begin{subfigure}[t]{0.5\textwidth}
		\centering
		\includegraphics[width=0.90\textwidth]{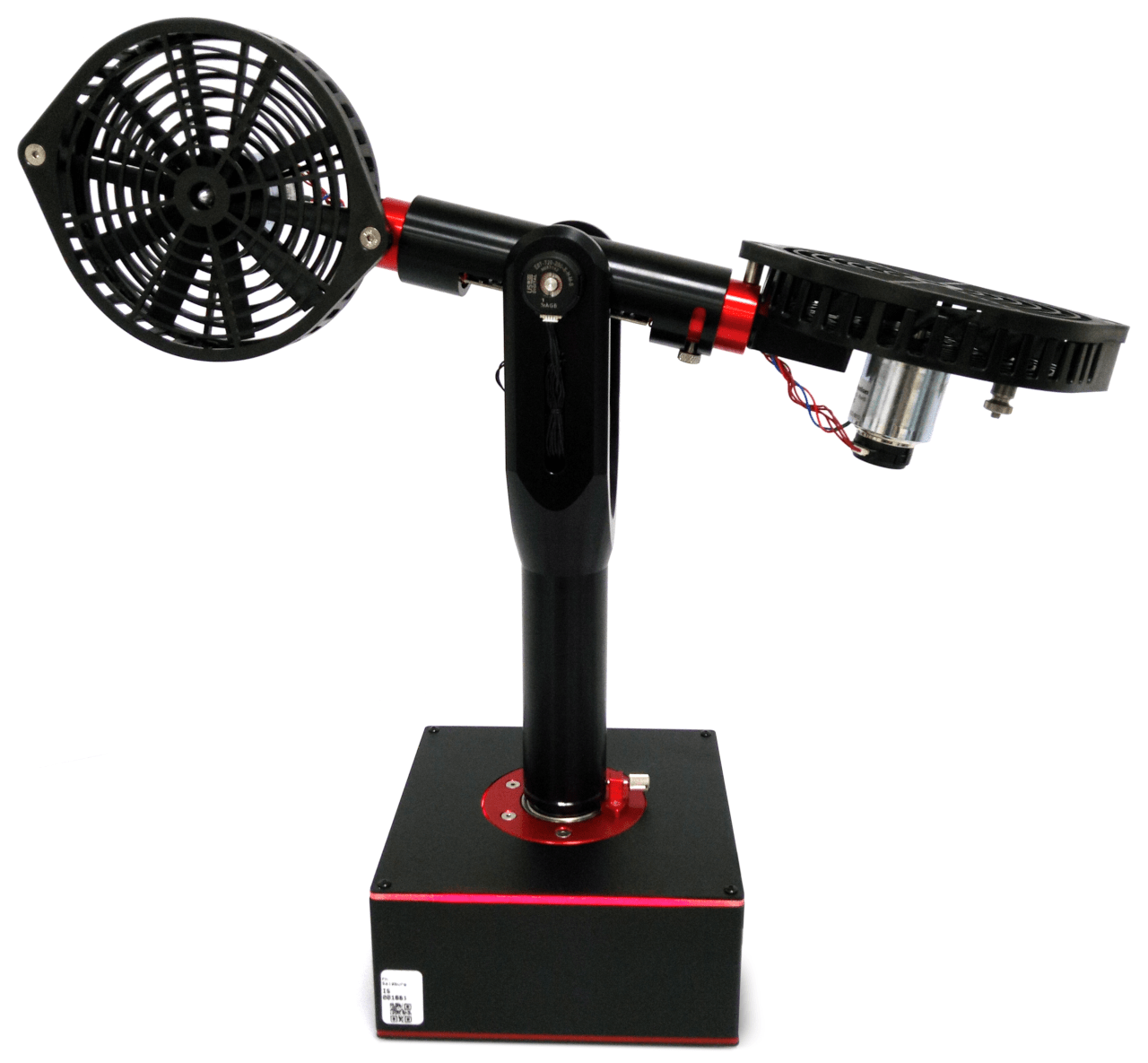}
	\end{subfigure}%
	~
	\begin{subfigure}[t]{0.5\textwidth}
		\centering
		\tikzset{every picture/.style={line width=0.75pt}} %set default line width to 0.75pt        

\begin{tikzpicture}[x=0.75pt,y=0.75pt,yscale=-1,xscale=1]
	
	%Shape: Can [id:dp2807595746319207] 
	\draw  [fill={rgb, 255:red, 155; green, 155; blue, 155 }  ,fill opacity=1 ] (256.04,68.5) -- (256.04,86.3) .. controls (256.04,87.9) and (251.71,89.2) .. (246.38,89.2) .. controls (241.04,89.2) and (236.71,87.9) .. (236.71,86.3) -- (236.71,68.5) .. controls (236.71,66.9) and (241.04,65.6) .. (246.38,65.6) .. controls (251.71,65.6) and (256.04,66.9) .. (256.04,68.5) .. controls (256.04,70.1) and (251.71,71.4) .. (246.38,71.4) .. controls (241.04,71.4) and (236.71,70.1) .. (236.71,68.5) ;
	%Shape: Can [id:dp6353282186342919] 
	\draw  [fill={rgb, 255:red, 195; green, 195; blue, 195 }  ,fill opacity=1 ] (210.88,55.71) -- (210.88,66.48) .. controls (210.88,69.45) and (226.32,71.86) .. (245.38,71.86) .. controls (264.43,71.86) and (279.88,69.45) .. (279.88,66.48) -- (279.88,55.71) .. controls (279.88,52.74) and (264.43,50.33) .. (245.38,50.33) .. controls (226.32,50.33) and (210.88,52.74) .. (210.88,55.71) .. controls (210.88,58.69) and (226.32,61.1) .. (245.38,61.1) .. controls (264.43,61.1) and (279.88,58.69) .. (279.88,55.71) ;
	%Shape: Circle [id:dp7347879821876429] 
	\draw  [fill={rgb, 255:red, 255; green, 255; blue, 255 }  ,fill opacity=1 ] (181.62,71.93) .. controls (181.62,67.54) and (178.06,63.98) .. (173.67,63.98) .. controls (169.28,63.98) and (165.72,67.54) .. (165.72,71.93) .. controls (165.72,76.32) and (169.28,79.88) .. (173.67,79.88) .. controls (178.06,79.88) and (181.62,76.32) .. (181.62,71.93) -- cycle ;
	%Shape: Arc [id:dp7323970907311912] 
	\draw  [draw opacity=0][fill={rgb, 255:red, 217; green, 217; blue, 217 }  ,fill opacity=1 ] (140.62,62.19) .. controls (140.63,61.84) and (140.64,61.49) .. (140.64,61.14) .. controls (140.64,43.28) and (124.98,28.8) .. (105.66,28.8) .. controls (86.34,28.8) and (70.68,43.28) .. (70.68,61.14) .. controls (70.68,61.14) and (70.68,61.15) .. (70.68,61.15) -- (105.66,61.14) -- cycle ; \draw   (140.62,62.19) .. controls (140.63,61.84) and (140.64,61.49) .. (140.64,61.14) .. controls (140.64,43.28) and (124.98,28.8) .. (105.66,28.8) .. controls (86.34,28.8) and (70.68,43.28) .. (70.68,61.14) .. controls (70.68,61.14) and (70.68,61.15) .. (70.68,61.15) ;  
	%Straight Lines [id:da03552957667734402] 
	\draw    (140.67,59.73) -- (140.62,66.54) ;
	%Straight Lines [id:da7131864785630844] 
	\draw    (70.68,61.15) -- (70.62,66.54) ;
	%Straight Lines [id:da5083741193229165] 
	\draw    (181.17,52.45) -- (181.62,71.93) ;
	%Straight Lines [id:da34952969256724487] 
	\draw    (165.39,51.81) -- (165.72,71.93) ;
	%Straight Lines [id:da7939609503204366] 
	\draw    (210.44,61.63) -- (188.03,61.63) ;
	%Straight Lines [id:da10210205710637821] 
	\draw    (157.63,61.88) -- (140.44,61.88) ;
	%Shape: Ellipse [id:dp03136538372638298] 
	\draw  [fill={rgb, 255:red, 222; green, 222; blue, 222 }  ,fill opacity=1 ] (140.62,66.54) .. controls (140.62,47.66) and (124.95,32.35) .. (105.62,32.35) .. controls (86.29,32.35) and (70.62,47.66) .. (70.62,66.54) .. controls (70.62,85.42) and (86.29,100.72) .. (105.62,100.72) .. controls (124.95,100.72) and (140.62,85.42) .. (140.62,66.54) -- cycle ;
	%Shape: Rectangle [id:dp15261297448268296] 
	\draw  [fill={rgb, 255:red, 207; green, 207; blue, 207 }  ,fill opacity=1 ] (138.43,159) -- (208.57,159) -- (208.57,177.29) -- (138.43,177.29) -- cycle ;
	%Shape: Rectangle [id:dp12788171569056672] 
	\draw  [fill={rgb, 255:red, 155; green, 155; blue, 155 }  ,fill opacity=1 ] (138.43,177.29) -- (208.43,177.29) -- (208.43,197) -- (138.43,197) -- cycle ;
	%Shape: Can [id:dp23050029456163146] 
	\draw  [fill={rgb, 255:red, 255; green, 255; blue, 255 }  ,fill opacity=1 ] (180.33,129.87) -- (180.33,170.07) .. controls (180.33,171.26) and (177.12,172.22) .. (173.17,172.22) .. controls (169.21,172.22) and (166,171.26) .. (166,170.07) -- (166,129.87) .. controls (166,128.68) and (169.21,127.72) .. (173.17,127.72) .. controls (177.12,127.72) and (180.33,128.68) .. (180.33,129.87) .. controls (180.33,131.06) and (177.12,132.02) .. (173.17,132.02) .. controls (169.21,132.02) and (166,131.06) .. (166,129.87) ;
	%Straight Lines [id:da6363762518243641] 
	\draw    (173.67,79.88) -- (173.33,130.33) ;
	%Shape: Ellipse [id:dp7980175556337241] 
	\draw  [fill={rgb, 255:red, 128; green, 128; blue, 128 }  ,fill opacity=1 ] (274.45,55.43) .. controls (274.45,53.61) and (261.43,52.14) .. (245.38,52.14) .. controls (229.32,52.14) and (216.3,53.61) .. (216.3,55.43) .. controls (216.3,57.24) and (229.32,58.71) .. (245.38,58.71) .. controls (261.43,58.71) and (274.45,57.24) .. (274.45,55.43) -- cycle ;
	%Shape: Ellipse [id:dp30525487527152784] 
	\draw  [fill={rgb, 255:red, 155; green, 155; blue, 155 }  ,fill opacity=1 ] (136.21,67) .. controls (136.21,50.04) and (122.46,36.29) .. (105.49,36.29) .. controls (88.53,36.29) and (74.78,50.04) .. (74.78,67) .. controls (74.78,83.96) and (88.53,97.71) .. (105.49,97.71) .. controls (122.46,97.71) and (136.21,83.96) .. (136.21,67) -- cycle ;
	%Shape: Rectangle [id:dp05890130113315384] 
	\draw   (188.26,49.6) -- (158.11,49.6) -- (158.11,74.4) -- (188.26,74.4) -- cycle ;
	%Shape: Arc [id:dp4469434392811238] 
	\draw  [draw opacity=0][fill={rgb, 255:red, 255; green, 255; blue, 255 }  ,fill opacity=1 ] (165.39,51.81) .. controls (166.34,48.98) and (169.47,46.89) .. (173.2,46.89) .. controls (177.19,46.89) and (180.51,49.29) .. (181.17,52.45) -- (173.2,53.62) -- cycle ; \draw   (165.39,51.81) .. controls (166.34,48.98) and (169.47,46.89) .. (173.2,46.89) .. controls (177.19,46.89) and (180.51,49.29) .. (181.17,52.45) ;  
	%Straight Lines [id:da7802877501688505] 
	\draw    (173.67,74.4) -- (173.67,71.93) ;
	%Shape: Pie [id:dp0949879173805408] 
	\draw  [fill={rgb, 255:red, 74; green, 74; blue, 74 }  ,fill opacity=1 ] (112.95,96.57) .. controls (108.55,97.72) and (103.98,97.88) .. (99.59,97.07) -- (105.29,66.97) -- cycle ;
	%Shape: Pie [id:dp7768891998446639] 
	\draw  [fill={rgb, 255:red, 74; green, 74; blue, 74 }  ,fill opacity=1 ] (87.44,91.68) .. controls (83.76,89.01) and (80.75,85.57) .. (78.59,81.66) -- (105.49,67) -- cycle ;
	%Shape: Pie [id:dp7786792541292846] 
	\draw  [fill={rgb, 255:red, 74; green, 74; blue, 74 }  ,fill opacity=1 ] (74.92,67.61) .. controls (74.82,63.06) and (75.72,58.59) .. (77.52,54.49) -- (105.49,67) -- cycle ;
	%Shape: Pie [id:dp539271339986428] 
	\draw  [fill={rgb, 255:red, 74; green, 74; blue, 74 }  ,fill opacity=1 ] (85.66,43.73) .. controls (89.1,40.77) and (93.14,38.63) .. (97.45,37.44) -- (105.49,67) -- cycle ;
	%Shape: Pie [id:dp6177285530083807] 
	\draw  [fill={rgb, 255:red, 74; green, 74; blue, 74 }  ,fill opacity=1 ] (111.66,37.06) .. controls (116.11,37.99) and (120.26,39.89) .. (123.84,42.58) -- (105.29,66.97) -- cycle ;
	%Shape: Pie [id:dp6708399575634043] 
	\draw  [fill={rgb, 255:red, 74; green, 74; blue, 74 }  ,fill opacity=1 ] (133.21,53.97) .. controls (135.17,58.07) and (136.18,62.53) .. (136.21,67) -- (105.57,67.04) -- cycle ;
	%Shape: Pie [id:dp6763152595335264] 
	\draw  [fill={rgb, 255:red, 74; green, 74; blue, 74 }  ,fill opacity=1 ] (132.95,80.11) .. controls (131.02,84.22) and (128.2,87.82) .. (124.77,90.68) -- (105.32,67) -- cycle ;
	%Shape: Circle [id:dp02940363867904483] 
	\draw  [fill={rgb, 255:red, 74; green, 74; blue, 74 }  ,fill opacity=1 ] (97.62,66.97) .. controls (97.62,62.73) and (101.06,59.3) .. (105.29,59.3) .. controls (109.52,59.3) and (112.96,62.73) .. (112.96,66.97) .. controls (112.96,71.2) and (109.52,74.63) .. (105.29,74.63) .. controls (101.06,74.63) and (97.62,71.2) .. (97.62,66.97) -- cycle ;
	%Shape: Ellipse [id:dp40587400220031866] 
	\draw  [fill={rgb, 255:red, 74; green, 74; blue, 74 }  ,fill opacity=1 ] (237.79,55.43) .. controls (237.79,54.27) and (241.18,53.33) .. (245.38,53.33) .. controls (249.57,53.33) and (252.96,54.27) .. (252.96,55.43) .. controls (252.96,56.59) and (249.57,57.53) .. (245.38,57.53) .. controls (241.18,57.53) and (237.79,56.59) .. (237.79,55.43) -- cycle ;
	%Curve Lines [id:da680978482216529] 
	\draw [color={rgb, 255:red, 255; green, 255; blue, 255 }  ,draw opacity=1 ][line width=1.5]    (162.14,136.14) .. controls (139,141.57) and (168.14,159.57) .. (198.71,145.29) ;
	%Curve Lines [id:da5899842367986006] 
	\draw    (162.14,136.14) .. controls (139.69,141.41) and (166.44,158.5) .. (195.97,146.49) ;
	\draw [shift={(198.71,145.29)}, rotate = 154.95] [fill={rgb, 255:red, 0; green, 0; blue, 0 }  ][line width=0.08]  [draw opacity=0] (10.72,-5.15) -- (0,0) -- (10.72,5.15) -- (7.12,0) -- cycle    ;
	%Curve Lines [id:da055838222223906664] 
	\draw [color={rgb, 255:red, 255; green, 255; blue, 255 }  ,draw opacity=1 ][line width=1.5]    (176.43,83) .. controls (188.5,83.4) and (196.41,72.81) .. (197.23,64.6) .. controls (198.05,56.39) and (191.5,42.88) .. (174.8,43.42) ;
	%Curve Lines [id:da6958782099635357] 
	\draw    (176.43,83) .. controls (188.5,83.4) and (196.41,72.81) .. (197.23,64.6) .. controls (198,56.88) and (192.25,44.49) .. (177.69,43.47) ;
	\draw [shift={(174.8,43.42)}, rotate = 358.17] [fill={rgb, 255:red, 0; green, 0; blue, 0 }  ][line width=0.08]  [draw opacity=0] (10.72,-5.15) -- (0,0) -- (10.72,5.15) -- (7.12,0) -- cycle    ;
	
	% Text Node
	\draw (227.61,102.86) node [anchor=north west][inner sep=0.75pt]   [align=left] {Fan 0};
	% Text Node
	\draw (86.87,102.86) node [anchor=north west][inner sep=0.75pt]   [align=left] {Fan 1};
	% Text Node
	\draw (187,126) node [anchor=north west][inner sep=0.75pt]  [font=\normalsize] [align=left] {Yaw $\displaystyle \psi $};
	% Text Node
	\draw (194.09,33.2) node [anchor=north west][inner sep=0.75pt]   [align=left] {Pitch $\displaystyle \varTheta $};

\end{tikzpicture}
	\end{subfigure}
	\caption{The Quanser Aero~2~(left) and its schematic representation (right) in a 2-\ac{DoF} configuration.}
	\label{fig:cps}
\end{figure}

The \ac{CPS} in this work is represented by a demonstrator with two fans mounted on a balancing beam, manufactured by Quanser.
It has two \acp{DoF}, namely the pitch $\varPsi$ and the jaw $\varTheta$, as shown in \cref{fig:cps}. By controlling the airflow of two fans,
configurable target angles for both yaw and pitch can be maintained.

This demonstrator was selected for multiple reasons:
First, it provides system characteristics with a high dynamic range and small physical time constants, as typically found in
industrial discrete automation for piecewise goods production~\cite{candell_jr_industrial_2015}.
Secondly, when compared to a typical industrial robotic arm with 6 to 7 \acp{DoF}, the two \acp{DoF} make it a good starting candidate
to explore generative model-based approaches for honeypots.

\subsubsection*{System Description.}
The demonstrator is equipped with two actuators, specifically two DC-motors, each with a fan directly coupled to its output shaft.
The speed of each motor can be adjusted by varying the supply voltage between -24 to 24 volts.
An optical incremental encoder mounted on the opposite side of each DC-motor provides speed feedback.
Additional sensory input is provided by two incremental encoders, one on the pitch axis and the other on the jaw axis.
A control loop maintains a configurable target yaw and pitch by adjusting the airflow generated by the fans, implemented using 
a state-space-based \ac{LQR}.

\subsubsection*{OPC UA Information Model.}

\begin{figure}[ht]
	\centering
	\includegraphics[width=0.5\textwidth]{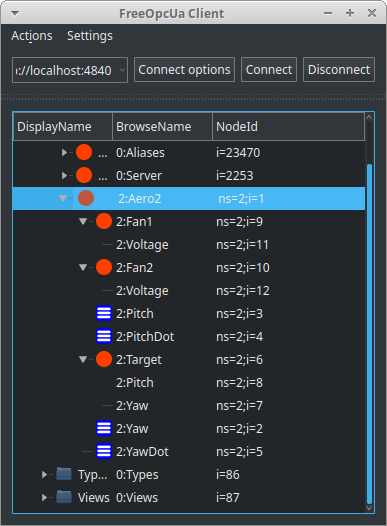}
	\caption{OPC UA Server Interface for the CPS queried with the \texttt{opcua-client-gui} \protect\footnotemark.}
	\label{fig:opcua}
\end{figure}

The \ac{opcua} information model provides an abstract definition framework for organizing and structuring data of a system.
A root node serves as the entry point to the model, with the objects node containing instances of various types.
In the context of the used \ac{CPS}, the \enquote{Fan}-type and \enquote{Target}-type are defined within this model, as shown in \cref{fig:opcua}.
While the \enquote{Fan}-type is solely set readable, the \enquote{Target}-type supports read-write access.

\footnotetext{\url{https://github.com/FreeOpcUa/opcua-client-gui} }

\section{Dataset}\label{sec:data}
\begin{figure}[ht]
	\centering
	\includegraphics[width=0.99\textwidth]{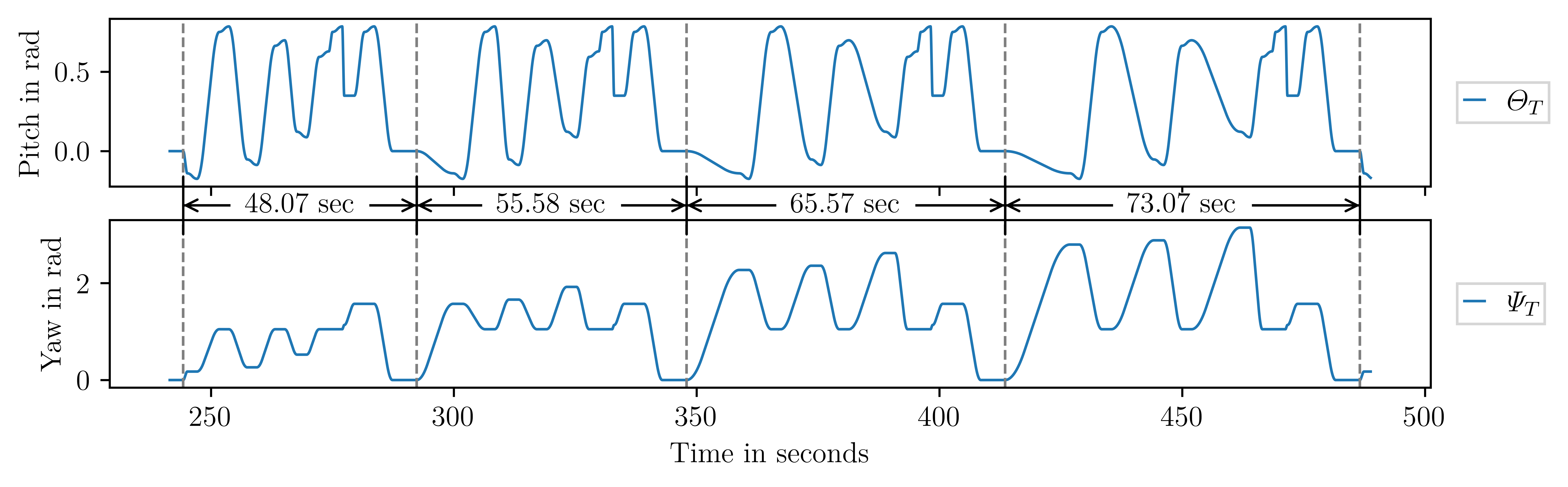}
	\caption{The target yaw $\varPsi_T$ and pitch $\varTheta_T$ angles of the CPS over time, realizing four sequences repeated in the same order multiple times. 
	 The duration of each sequence is annotated in seconds.}
    \label{fig:dataset_cycles}
\end{figure}

Using the \ac{CPS} introduced in \cref{sec:cps} a dataset is created.
By commanding target yaw $\varPsi_T$ and pitch $\varTheta_T$ angles over time, different device poses are repeatedly realized.
The resulting cyclic process resembles a pick-and-place operation. The cyclic process itself consists of four different sequences, as visualized in \cref{fig:dataset_cycles},
which are repeatedly being carried out. The duration of each sequence is as well given in \cref{fig:dataset_cycles}. 

\begin{figure}[h!]
	\centering
	\includegraphics[width=0.99\textwidth]{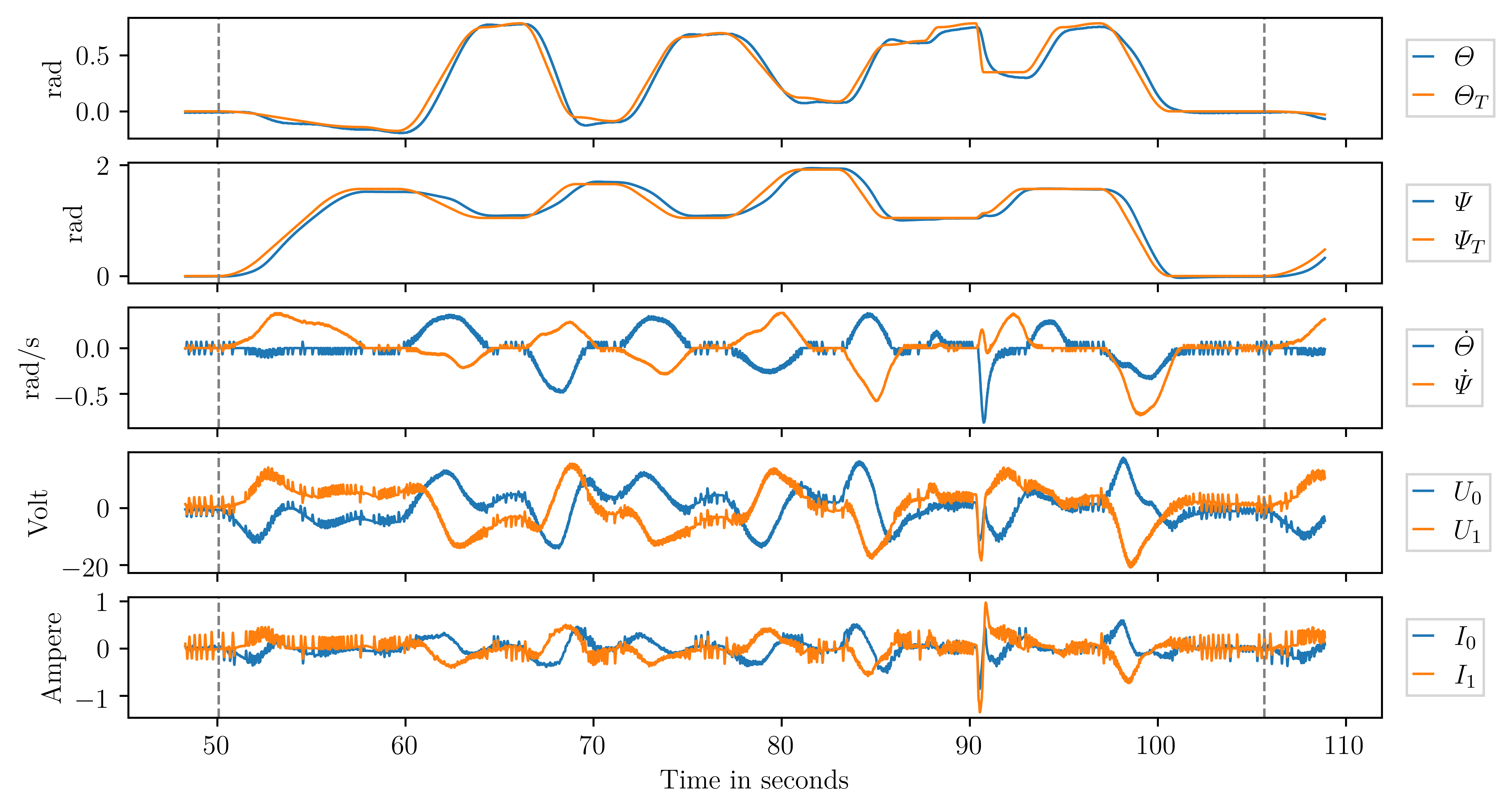}
	\caption{The second sequence of the cyclic process with start and end marked by dashed vertical lines. Showing motor voltages ($U_0$, $U_1$), currents ($I_0$, $I_1$), angular velocities ($\dot{\varTheta}$, $\dot{\varPsi}$), actual pitch and yaw ($\varTheta$, $\varPsi$), and target pitch and yaw ($\varTheta_T$, $\varPsi_T$). The actual pitch and yaw are slightly dragging behind their targets.} 
	\label{fig:dataset_cycle2}
\end{figure}

While running the cyclic process, data is sampled at a rate of 500 Hz over a total duration of 2 hours. 
The resulting dataset is in CSV-format and stores twelve different variables \footnote{The dataset is available under \url{https://www.github.com/JRC-ISIA/paper-2024-eurocast-honeypot} }.
These variables include motor voltages ($U_0$, $U_1$), actual yaw and pitch angles ($\varTheta$, $\varPsi$),
actual yaw and pitch angular velocities ($\dot{\varTheta}$, $\dot{\varPsi}$), and target yaw and pitch angles ($\varTheta_T$, $\varPsi_T$).
A detailed listing of each variable is given in \cref{tab:data} while in \cref{fig:dataset_cycle2} selected trajectories of the second sequence are shown.

\begin{table}[H]
	\centering
	\caption{Description of variables in the published CSV data file.}
	\begin{tabular}{llllcc}
		\toprule
		\textbf{Col.} & \textbf{Name} & \textbf{Description}                 & \textbf{Type} & \textbf{Symbol}   & \textbf{Unit} \\
		\midrule
		0             & Time          & Elapsed time since measurement start &   --            & $t$               & s             \\
		1             & Voltage0      & DC-motor 0 voltage            & output        & $U_0$             & V             \\
		2             & Voltage1      & DC-motor 1 voltage                                  & output        & $U_1$             & V             \\
		3             & Current0      & DC-motor 0 current            & measured      & $I_0$             & A             \\
		4             & Current1      & DC-motor 1 current                                   & measured      & $I_1$             & A             \\
		5             & MotorSpeed0   & Rotational speed of fan 0     & measured      & $s_0$             & rpm           \\
		6             & MotorSpeed1   & Rotational speed of fan 1                                   & measured      & $s_1$             & rpm           \\
		7             & Yaw           & Actual yaw angle                     & measured      & $\varPsi$         & rad           \\
		8             & Pitch         & Actual pitch angle                   & measured      & $\varTheta$       & rad           \\
		9             & TargetYaw     & Target yaw angle                     & input         & $\varPsi_T$       & rad           \\
		10            & TargetPitch   & Target pitch angle                   & input         & $\varTheta_T$     & rad           \\
		11            & YawDot        & Yaw, angular velocity                & estimated     & $\dot{\varPsi}$   & rad/s         \\
		12            & PitchDot      & Pitch, angular velocity              & estimated     & $\dot{\varTheta}$ & rad/s         \\
		\bottomrule
	\end{tabular}
	\label{tab:data}
\end{table}

\section{Generative Model}\label{sec:model}

In the machine-learning domain, the relevant subdomain is time-series forecasting or prediction.
For this particular application, multiple variables over a longer time period are to be predicted,
a task known as multivariate multi-step forecasting.

Different forecasting strategies can be used for multi-step forecasts,
most commonly used is a recursive strategy with single-step forecasts or multi-step forecasts~\cite{aufaure_machine_2013}.
As in our experiments, single-step forecasts resulted in a poor performance with multiple recursions,
it was opted to use multi-step forecasts. In the end a look-back of 4 seconds and look-ahead of 0.4 seconds is used,
as such the generative model uses the last 4 seconds (2000 samples) to generate the next 0.4 seconds (200 samples).
To reduce the complexity for this proof-of-concept, current and speed of the DC-motors are not replicated, 
resulting in 8 variables, of which trajectories are to be generated.
For data normalization a min-max scaler was used for all 8 variables.

\begin{figure}[ht]
	\centering
	\includegraphics[width=0.99\textwidth]{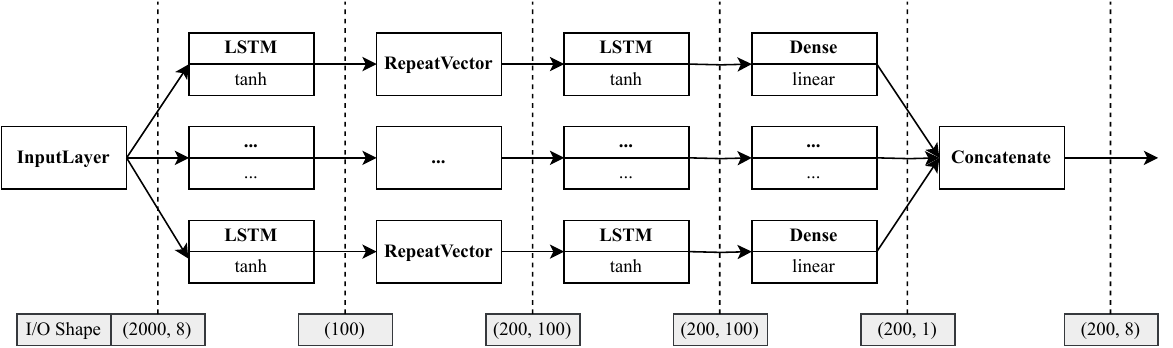}
	\caption{Structure of the employed generative model: individually trained ED-LSTM, each for a single variable, are concatenated to build a multi-variate generative model.}
	\label{fig:model}
\end{figure}

For this proof-of-concept, the focus is put on \ac{LSTM}, as they are reported
to be overall well suited for prediction tasks~\cite{chandra_evaluation_2021}.
The specific subtype employed is known as Encoder-Decoder \acp{LSTM}. In initial experiments  a single ED-LSTM
with the final decoder-stage outputting all 8 variables was used. This however turned out to be challenging to train.
Instead, the final approach involves using multiple smaller single-output models, which are trained individually.
This method required less tuning and provides more stable results.
Overall, the training is performed using the Adam optimizer with a default learning rate of 1e-3 and a default \ac{MSE} loss function.
As shown in \cref{fig:model},
the individually trained models are joined, outputting the multi-step forecasts for the 8 variables.
As such, one resulting large model can be deployed to the GPU.

\begin{figure}
	\centering
	\includegraphics[width=0.99\textwidth]{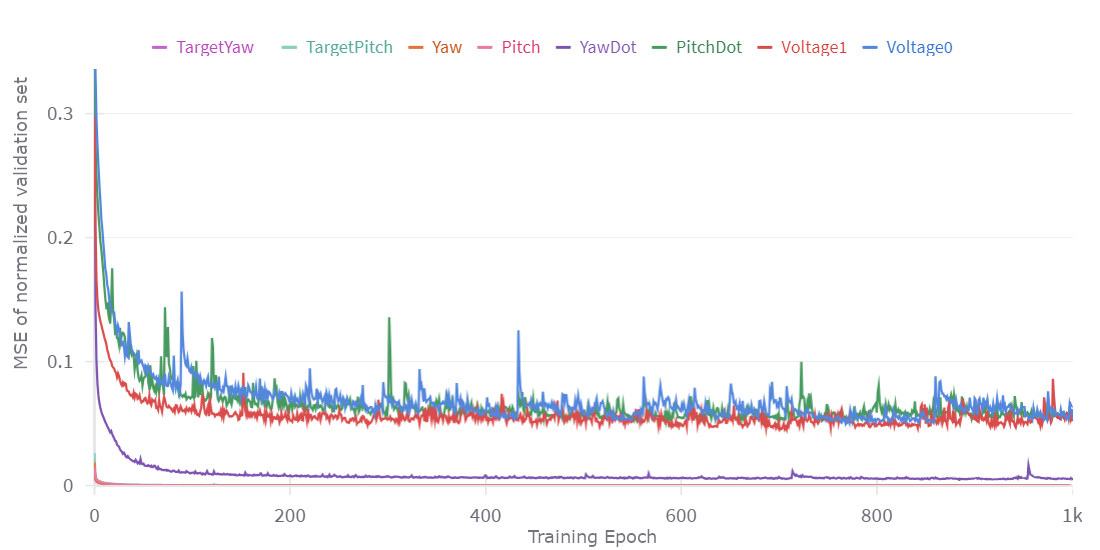} 
	\caption{Training of the generative model over 1000 epochs.}
	\label{fig:train}
\end{figure}

In \cref{fig:train} the training progress of the model is visualized, showing the \ac{MSE} over training epochs.
Variables with higher frequency components, such as Voltage and Angular Velocities, are more challenging.
These high-frequency components are harder to model accurately, 
which results in a slower reduction of the \ac{MSE} and higher final \ac{MSE} values compared to variables with lower frequency spectra.

\subsection{Implementation}\label{sec:impl}

With the generative model defined in \cref{sec:model} , the next step is to implement it as part of an \ac{opcua} server.
The implementation is Python-based and follows the Producer/Consumer Pattern, resulting in two CPU threads working in parallel.

\begin{figure}[ht]
	\centering
	\includegraphics[width=0.7\textwidth]{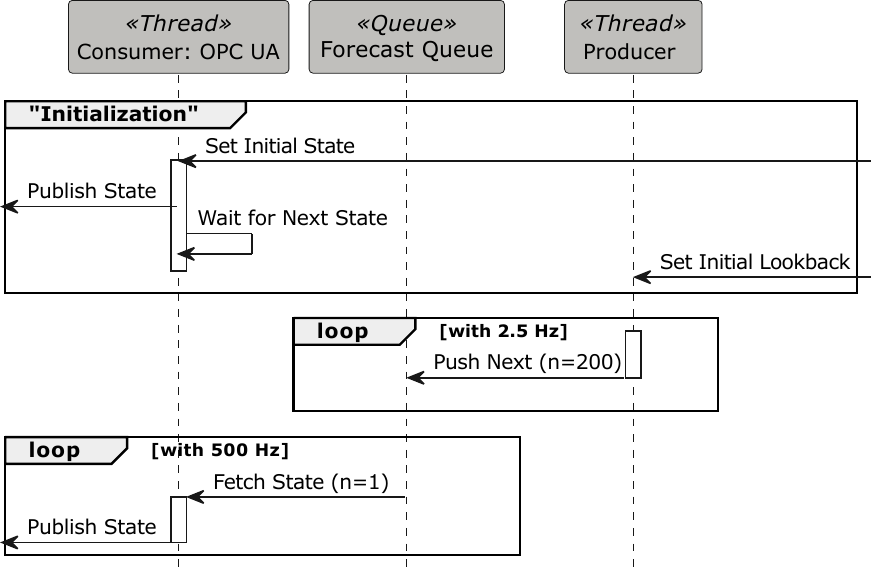}
	\caption{Sequence diagram of the honeypot implementation.}
	\label{fig:seqdia}
\end{figure}

\Cref{fig:seqdia} provides an overview of the implementation.
In the initialization phase, the initial look-back and state for the variables is received,
either directly from the real \ac{CPS} or via an aggregation service as described in \cite{hirsch_opc_2023}.
Afterwards, the producer thread oversees the execution of the generative model, which itself is deployed to a GPU.
Once a single forecast is finished, a batch of new values is appended to a thread-safe queue at a 2.5 Hz rate.
Meanwhile, the consumer thread operates the \ac{opcua} server,
publishing updates of the variables at a 500 Hz rate.
By doing so, it removes the values from the thread-safe queue one by one.

\section{Results}\label{sec:results}

This section summarizes the findings from our experiments regarding the evaluation of 
the proposed generative model (see \cref{sec:model}) with the newly created dataset (see \cref{sec:data}).
Besides the model performance metrics, the required hardware resources are noted as well.

\subsection{Performance Evaluation}

\begin{figure}[ht]
	\centering
	\includegraphics[width=0.99\textwidth]{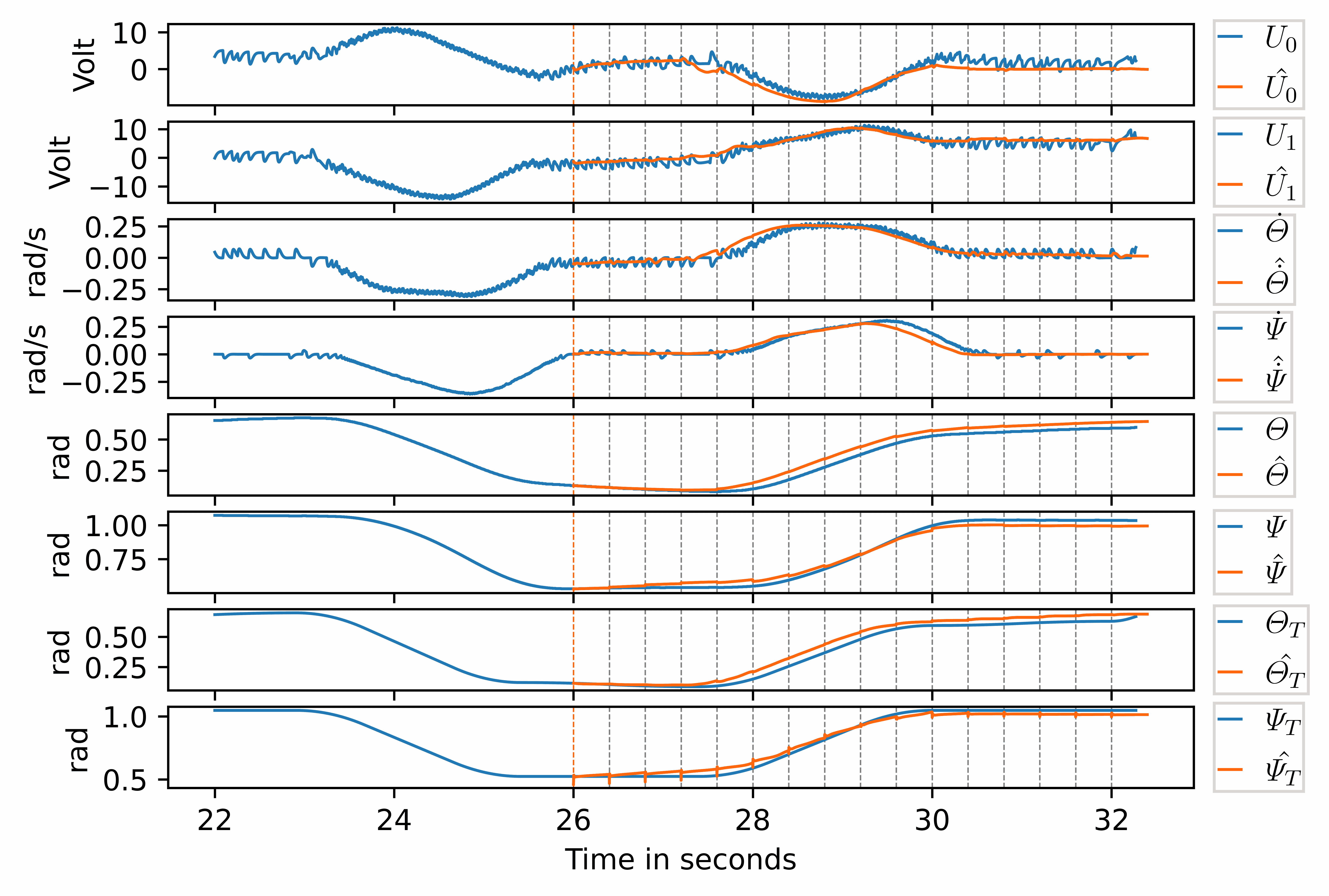}
	\caption{The initial four seconds of data are used as input for the generative model, with a forecast of 6.4 seconds (16 steps) representing the state space of the trajectory. Each computation step is indicated by a vertical dashed line.}
	\label{fig:forecast16}
\end{figure}

The performance of the generative model is evaluated by comparing 
the generated trajectories to existing trajectories from the dataset.
\cref{fig:forecast16} shows a forecast of 6.4 seconds,
with the time axis starting at 22 seconds. Up to the time of 26 seconds,
the trajectories are used as input (look-back) for the generative model.
Based on this input, the next 0.4 seconds are produced (look-ahead).
Each look-ahead segment is indicated by a vertical dashed line in the figure.
Since the generative model is operating isolated from the real \ac{CPS},
every look-ahead segment becomes part of the next look-back for the generative model.
Therefore, in \cref{fig:forecast16}, at a time of 30 seconds,
the input for the generative model solely consists of previously generated segments.
Ideally, this process can continue indefinitely, with the model generating trajectories based on a single initial input.

Following the  target pitch trajectory $\varPsi_T$ in \cref{fig:forecast16} more closely,
discontinuities are evident in the estimated trajectory at the beginning of each look-ahead segment.
Regarding ripple, while observed in the trajectories for the voltages $U_0$ and $U_1$, 
the replicated trajectories notably lack these features.
In similar fashion, this holds true for the trajectories of the angular velocities $\dot{\varPsi}$ and $\dot{\varTheta}$.
However it must be noted, that compared to $\dot{\varTheta}$, only a lower ripple is present for $\dot{\varPsi}$ on the yaw axis.
There are multiple reasons for this. The lower joint, which allows for rotation around the yaw-axis, has a higher mechanical moment of inertia,
therefore better mechanically dampening higher-frequency components.
Additionally, the bearing of the lower joint is of a larger diameter, 
and it includes a slip-ring for electrical connections, together resulting in higher friction.
Combined, these mechanical constraints explain well the observed lower ripple for $\dot{\varPsi}$.
Referring back to \cref{fig:train}, the amount of ripple in the trajectories also correlates with the \ac{MSE} in the training results.

\begin{figure}[H]
	\centering
	\includegraphics[width=0.99\textwidth]{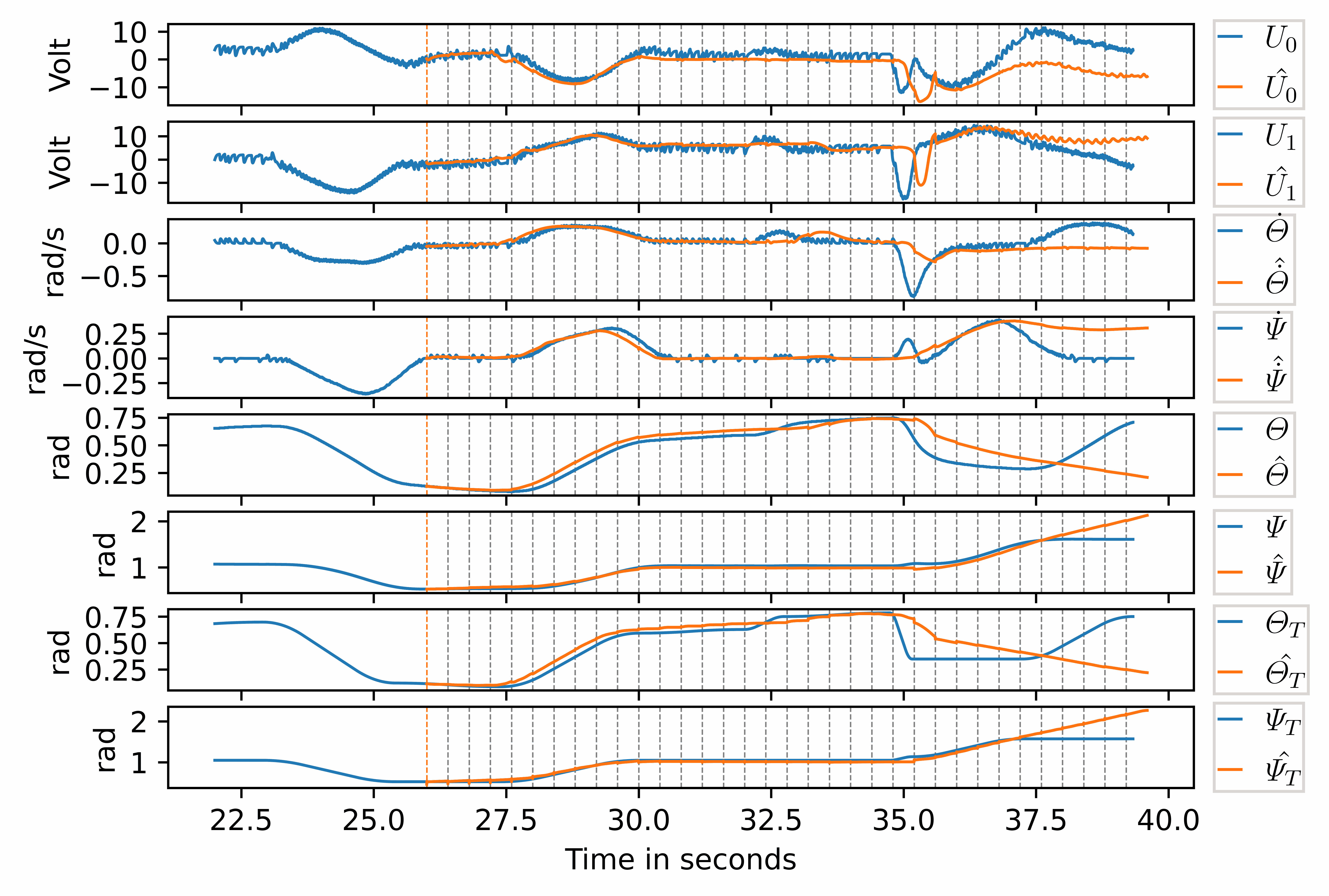}
	\caption{Similar to \cref{fig:forecast33}, but with a forecast of 13.2 seconds (33 steps).}	
	\label{fig:forecast33}
\end{figure}

A longer replicated trajectory of 13.2 seconds is shown in \cref{fig:forecast33}.
Due to the recursive forecast strategy, errors accumulate and the generated trajectories start to deviate.
For example, at 32 seconds, the generated trajectory is no longer in sync,
instead it is slightly shifted to the right on the time-axis.
This also can be observed at a time of 35 seconds for the voltages $U_0$ and $U_1$.
Towards the end around a time of 37 seconds in \cref{fig:forecast33}, 
all the generated trajectories clearly deviate, seemingly resulting in unstable drifting trajectories.

\begin{figure}[ht]
	\centering
	\includegraphics[width=0.99\textwidth]{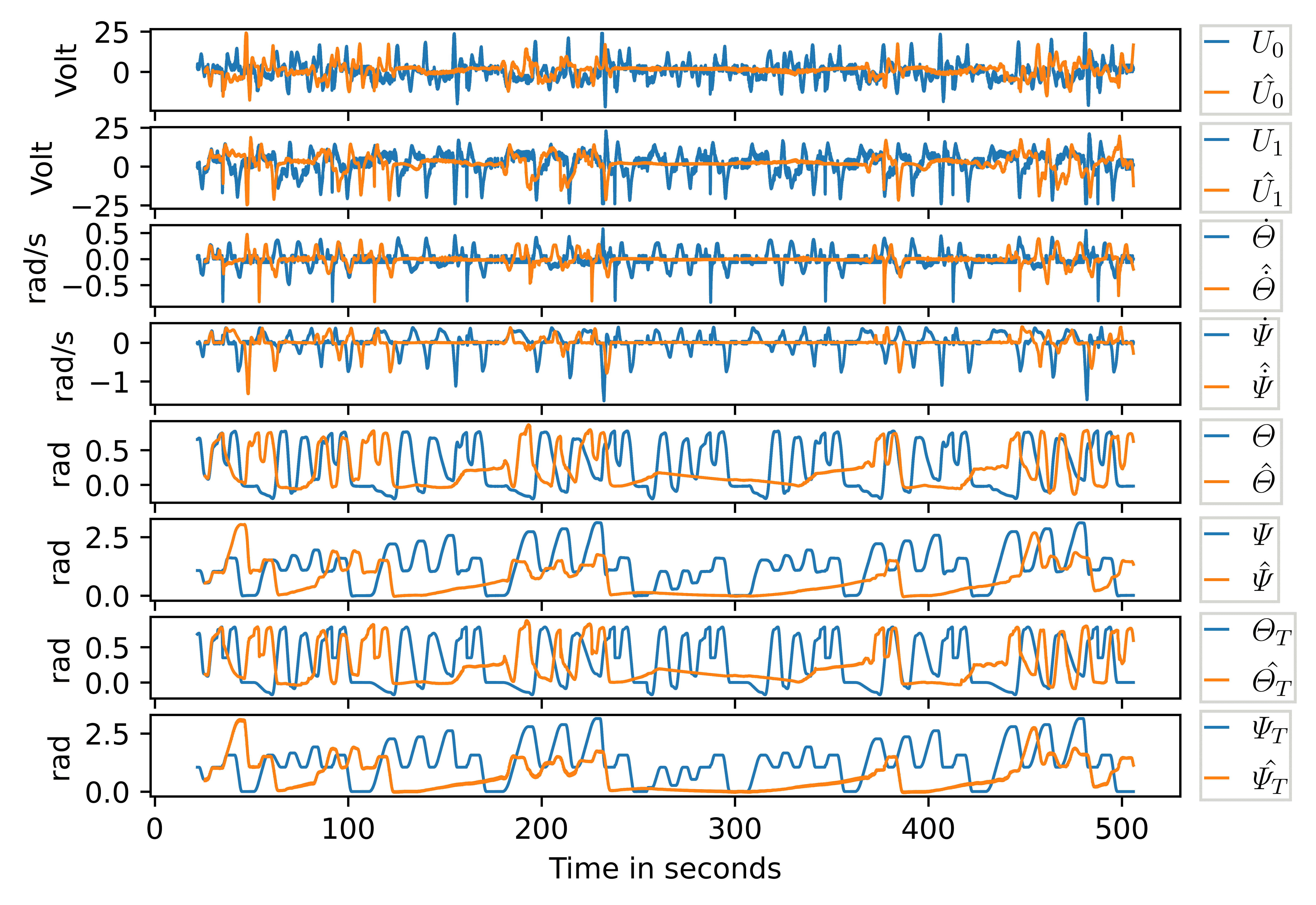}
	\caption{Long-term forecast of 8 Minutes (1200 Steps) representing the state space of the trajectory.}
	\label{fig:forecast1200}
\end{figure}

In \cref{fig:forecast1200}, the trajectories are replicated for over 8 minutes,
in total consisting of 1200 generated segments.
While clearly diverging from the original trajectories of the dataset, 
the generative model exhibits a seemingly stable behavior and
continues to generate patterns present in the original dataset with each variable remaining within its limit.

\begin{figure}[ht]
	\centering
	\includegraphics[width=0.99\textwidth]{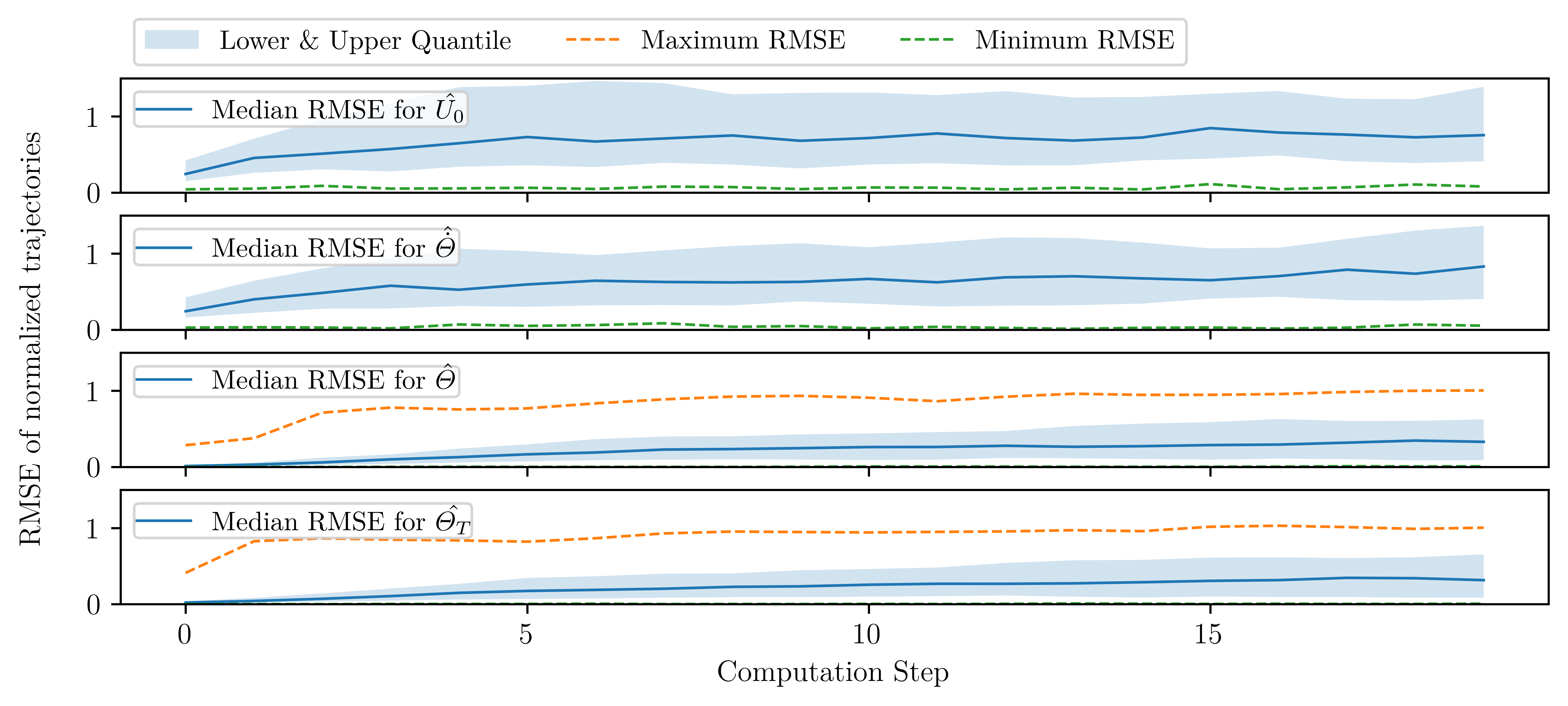}
	\caption{Root Mean Squared Error (\ac{RMSE}) of normalized values across computation steps of 301 trajectories. %for the motor voltage $\hat{U_0}$, angular velocity $\hat{\dot{\varTheta}}$, target pitch $\hat{\varTheta_T}$, and actual pitch $\hat{\varTheta}$.
		 The solid blue line indicates the median \ac{RMSE}, 
		and the light-blue shaded area represents the range between the lower and upper quantiles.
		The units on the y-axis are normalized based on the min-max scaling procedure of the ML model.}
	\label{fig:rmse}
\end{figure}
 
To evaluate the generative model statistically, 301 different trajectories are randomly selected from the validation dataset,
serving as initial look-back data for the generative model.
Consequently, for each input data, the model generated a trajectory consisting of 20 segments with eight variables, 
each made up of 200 data samples.
The error of each generated segment can be computed by comparing it to the actual trajectory using the \ac{RMSE}.
In the following, the distribution of the 301*20 obtained \ac{RMSE} values is examined to provide information regarding the stability of the generative model.
The results are shown in \cref{fig:rmse} for the motor voltage $\hat{U_0}$, angular velocity 
{\small $\hat{\dot{\varTheta}}$}, target pitch {\small $\hat{\varTheta_T}$}, and actual pitch {\small  $\hat{\varTheta}$} in this context.
The unit of the y-axis is normalized based on the min-max scaling procedure of the ML model to maintain a comparable magnitude for each of the target quantities.
In line with the observations of the sample trajectory shown in \cref{fig:forecast33}, the error of the generated trajectories generally accumulates over time.
Additionally, the large shaded areas representing the lower and upper quantiles indicate that the quality of the predictions strongly depends on the initially provided input data.
The proposed generative model is not able to generate trajectories equally well based on all initial look-back data.
For the voltages {\small $\hat{U_0}$} and {\small  $\hat{\dot{\varTheta}}$}, the results are less meaningful, as higher frequency components are missing in the replicated trajectories,
therefore causing a higher \ac{RMSE}.

\subsection{Required Hardware Resources}

\begin{figure}
	\centering
	\includegraphics[width=0.99\textwidth]{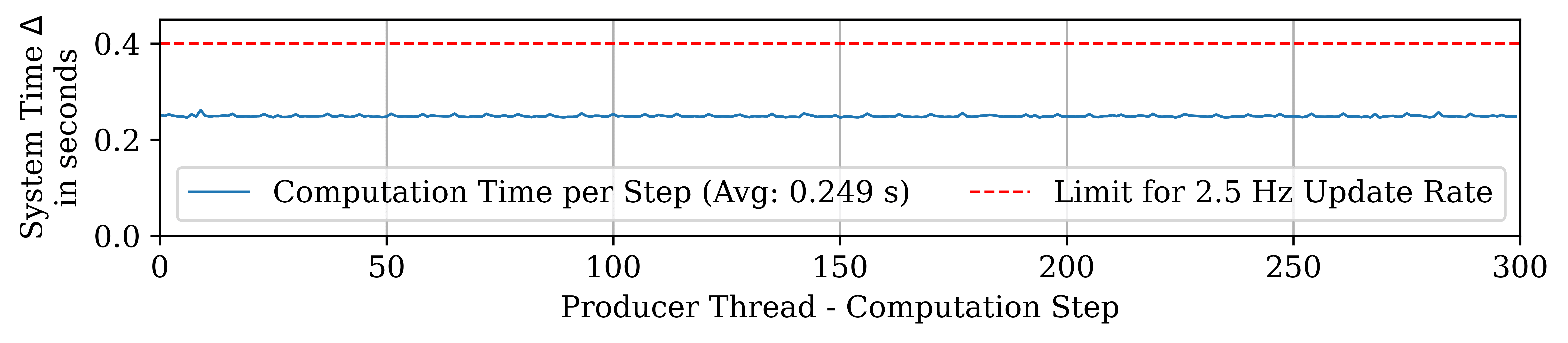}
	\caption{Required computation time of the producer thread monitored over $n=300$ steps.}
	\label{fig:step_time}
\end{figure}

The proposed implementation from \cref{sec:impl} is deployed on a relatively constrained system.
The system consists of an Intel Core i5 6400 series CPU,
which was released in 2015, and a low profile NVIDIA A2000 GPU with a power consumption of up to 75 W.
Performance measurements shown in \cref{fig:step_time} indicate that the minimum computation time for the producer thread is 0.244 seconds,
while the average computation time is about 0.250 seconds. The maximum computation time recorded is 0.261 seconds.
Therefore, the system's performance allows to maintain an update rate of 2.5 Hz for the producer thread,
enabling the consumer thread to publish updates at a rate of 500 Hz.

\section{Conclusion}\label{sec:conclusion}

This study demonstrates the successful deployment of an \ac{opcua} and generative model-based honeypot,
giving a first glimpse of its potential for cyber security applications.
The proposed implementation has managed a high sampling rate output,
as typically required for discrete automation.
The internal generative model is based on a \ac{LSTM} network, which exhibits interesting long-term forecasting characteristics.
While deviating from the actual trajectories over time, it maintains the trained variable ranges and continues to produce
plausible patterns, which were observed during training.
Besides the successful deployment of the generative model-based honeypot,
a dataset is presented, enabling further research in which the current results can serve as an initial baseline.

As for future work, several aspects need further attention.
The performance and scalability of different generative models need to be compared to improve the generative model,
explore different structures, and utilize physics-based optimizers and loss functions.
Modeling artificial ripple and noise is required. While it is not desirable for forecasts as it indicates over-fitting,
it is necessary for a plausible trajectory replication.
Then again, a highly accurate replication might not be desirable, a controllable degree of realism is required,
to find a balance to generate plausible trajectories without revealing sensitive information.
Also, focus can be put on the system understanding that the generative model acquires.
Ideally, the generative model will gain a deeper system understanding,
generalizing the training data, providing adequate responses for untrained scenarios.
Finally, the current implementation only allows the intruder to observe.
As such, providing a higher level of interaction for the intruder is a necessary enhancement.

\section*{Acknowledgment}

The financial support by the Christian Doppler Research Association, the
Austrian Federal Ministry for Digital and Economic Affairs and the Federal State
of Salzburg is gratefully acknowledged.

\bibliography{citations}

\end{document}